\documentclass{elsart}

\usepackage{amssymb}
\usepackage{epsfig}
\usepackage{amssymb}
\usepackage{epsf}
\usepackage{epsfig}
\usepackage{color}
\usepackage{ifpdf}
\usepackage{cite}

\newcommand{\ba}{\begin{eqnarray}}
\newcommand{\ea}{\end{eqnarray}}
\newcommand{\be}{\begin{equation}}
\newcommand{\ee}{\end{equation}}
\newcommand{\bd}{\begin{displaymath}}
\newcommand{\ed}{\end{displaymath}}
\newcommand{\azero}{a^{I=0}_{\pi\pi}}
\newcommand{\atwo}{a^{I=2}_{\pi\pi}}
\newcommand{\aI}{a^I_{\pi\pi}}
\newcommand{\E}{E^{I=2}_{\pi\pi}}
\newcommand{\dE}{\delta E^{I=2}_{\pi\pi}}
\newcommand{\lpipi}{l_{\pi\pi}^{I=2}}
\newcommand{\chipt}{$\chi$PT}
\newcommand{\Eq}[1]{Eq.~#1}
\newcommand{\Ref}[1]{Ref.~#1}
\newcommand{\Refs}[1]{Refs.~#1}
\newcommand{\Sect}[1]{Sect.~#1}
\newcommand{\gap}{\hspace{10pt}}
\newcommand{\Tab}[1]{Tab.~#1}
\newcommand{\Fig}[1]{Fig.~#1}
\newcommand{\fphy}{f_{\pi,\mathrm{phy}}}

\newcommand{\mev}{\mathrm{MeV}}
\newcommand{\gev}{\mathrm{GeV}}
\newcommand{\fm}{\mathrm{fm}}

\newcommand{\old}[1]{}
\newcommand{\new}[1]{#1}

\newcommand{\plotangle}{270}

\begin{document}

\begin{flushright}
  DESY 09-141\\
  SFB/CPP-09-82\\
  MS-TP-09-17
\end{flushright}

\begin{frontmatter}

\title{The $\pi^+\pi^+$ scattering length from\\maximally twisted mass lattice QCD}

\begin{center}
\includegraphics[width=100pt]{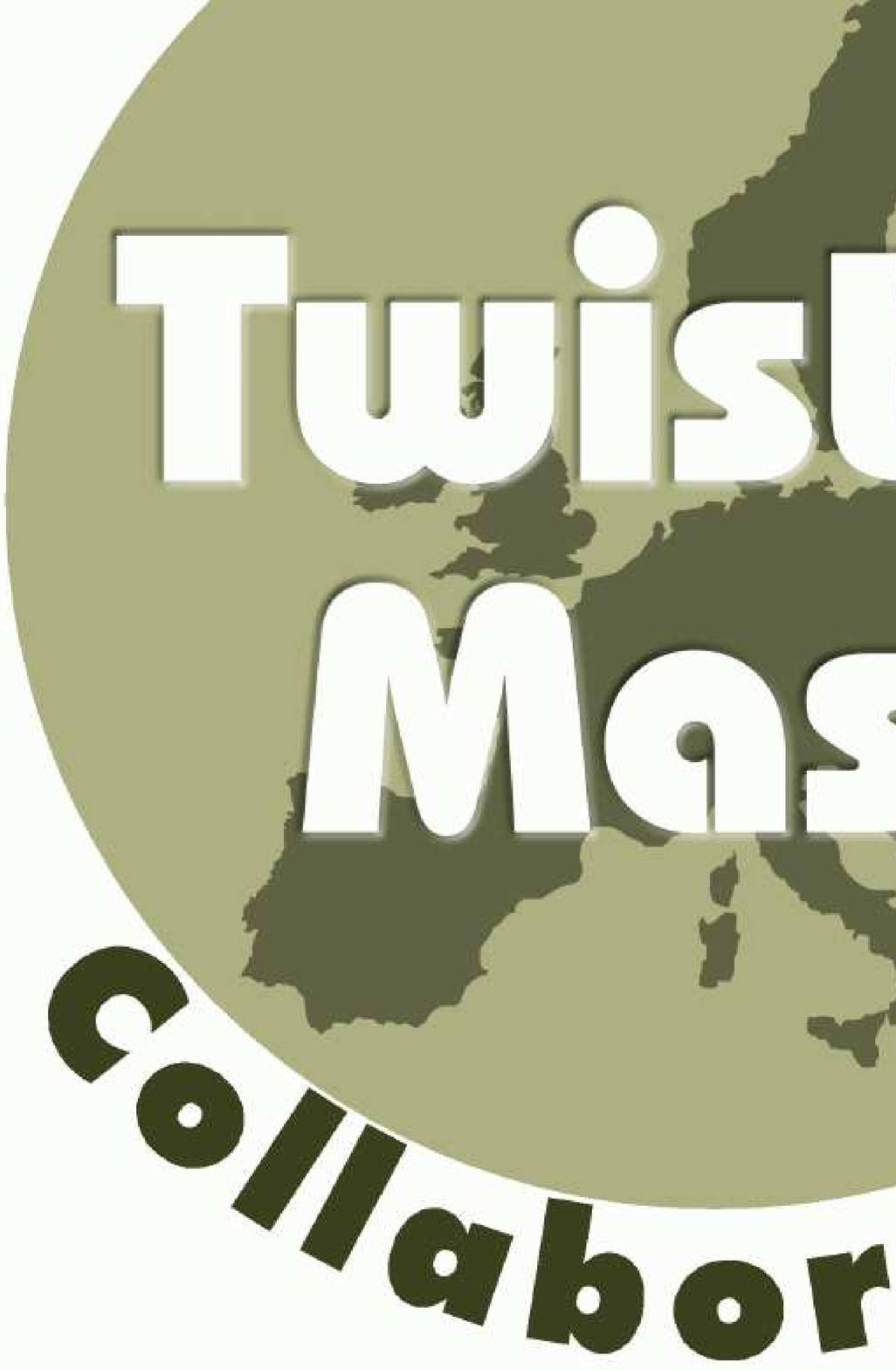}
\end{center}\vspace{10pt}

\author[DESY,Munster]{Xu Feng},
\author[DESY]{Karl Jansen},
\author[DESY]{Dru Renner}

\address[DESY]{NIC, DESY, Platanenallee 6, D-15738 Zeuthen, Germany}
\address[Munster]{Universit\"at M\"unster, Institut f\"ur Theoretische Physik,\\
Wilhelm-Klemm-Strasse 9, D-48149, Germany}

\begin{abstract}
We calculate the s-wave pion-pion scattering length in the isospin
$I=2$ channel in lattice QCD for pion masses ranging from $270~\mev$
to $485~\mev$ using two flavors of maximally twisted mass fermions
at a lattice spacing of $0.086~\fm$.  Additionally, we check for
lattice artifacts with one calculation at a finer lattice spacing of
$0.067~\fm$.  We use chiral perturbation theory at next-to-leading
order to extrapolate our results. At the physical pion mass, we find
$m_\pi \atwo=-0.04385\,(28)(38)$ for the scattering length,
where the first error is statistical and the second is our
estimate of several systematic effects.
\end{abstract}


\end{frontmatter}

\section{Introduction}

In the limit of massless up and down quarks, the resulting chiral
symmetry of QCD is spontaneously broken and consequently the meson
spectrum contains three massless Goldstone bosons, $\pi^\pm$ and
$\pi^0$.  Due to the unique role played by the pions, their
interactions are strongly determined by the underlying chiral
symmetry, and the s-wave pion-pion scattering lengths even vanish in
the chiral limit.  Thus the scattering lengths are sensitive to the
chiral dynamics of the strong interactions, and the non-perturbative
calculation thereof, the subject of this paper, is an integral part of
understanding the low energy properties of QCD.

In nature, the masses of the quarks are not zero but small and induce
an explicit but weak breaking of chiral symmetry.  Correspondingly,
the pions are not massless but light.  This breaking of chiral
symmetry is systematically treated in chiral perturbation theory
(\chipt) by considering the quark masses as perturbations.
Furthermore, the pion-pion scattering lengths no longer vanish and at
leading order (LO) in {\chipt} are predicted by
Weinberg~\cite{Weinberg:1966kf} solely in terms of the pion mass,
$m_\pi$, and the pion decay constant, $f_\pi$, as
\bd
m_\pi\azero\approx\frac{7m_\pi^2}{16\pi f_\pi^2}=0.160\,(1)
\gap\textmd{and}\gap m_\pi\atwo\approx-\frac{m_\pi^2}{8\pi
f_\pi^2}=-0.0456\,(1)\,,
\ed
where $\azero$ and $\atwo$ denote the isospin $I=0$ and $I=2$ s-wave
scattering lengths respectively.  The next-to-leading order (NLO)
corrections depend on unknown low energy constants, which can be
determined from experimental measurements or lattice calculations.
\old{However, a direct experimental measurement by E865 at
BNL~\cite{Pislak:2003sv} gives
\bd
m_\pi\azero=0.216\,(14) \gap\textmd{and}\gap
m_\pi\atwo=-0.0454\,(34)\,,
\ed
and a combination of experimental and theoretical inputs from
CGL~\cite{Colangelo:2000jc,Colangelo:2001df} produces a consistent
but more precise result of
\bd
m_\pi\azero=0.220\,(5) \gap\textmd{and}\gap
m_\pi\atwo=-0.0444\,(10)
\ed
for the scattering lengths.}

\new{The experimental measurement of $K^{\pm}\rightarrow
  \pi^+\pi^-e^{\pm}\nu$ ($K_{e4}$) decays by E865 at BNL~\cite{Pislak:2003sv} gives
\bd
m_\pi\azero=0.203\,(33) \gap\textmd{and}\gap
m_\pi\atwo=-0.055\,(23)\,.
\ed
When combined with constraints from {\chipt}, these measurments yield
\bd
m_\pi\azero=0.216\,(14) \gap\textmd{and}\gap
m_\pi\atwo=-0.0454\,(34)\,.
\ed
A combination of several experimental and theoretical inputs from
CGL~\cite{Colangelo:2000jc,Colangelo:2001df} produces a consistent but
more precise result of
\bd
m_\pi\azero=0.220\,(5) \gap\textmd{and}\gap
m_\pi\atwo=-0.0444\,(10)\,.
\ed
Additionally, the recent measurments of $K_{e4}$ decays~\cite{Batley:2007zz} and
$K^{\pm}\rightarrow\pi^{\pm}\pi^0\pi^0$ decays~\cite{Batley:2000zz} by NA48/2 at
CERN~\cite{NA48} give, without making any use of {\chipt} constraints,
\bd
m_\pi\azero=0.221\,(5) \gap\textmd{and}\gap
m_\pi\atwo=-0.0429\,(47).
\ed
Including {\chipt} in their analysis, NA48/2 finds~\cite{BBD}
\bd
m_\pi\azero=0.220\,(3) \gap\textmd{and}\gap
m_\pi\atwo=-0.0444\,(9).
\ed
The results are all consistent with each other and the most precise
results from NA48/2 are in agreement with the lattice results given
shortly.}

The non-perturbative determination of the scattering lengths from
lattice QCD is possible, despite the Euclidean nature of the
calculation, by using a finite size method due to
L\"uscher~\cite{Luscher:1985dn,Luscher:1986pf,Luscher:1990ck,Luscher:1990ux,Luscher:1991cf}.
This method capitalizes on the relationship between the energy
eigenvalues of a two pion system enclosed in a finite spatial box and
the scattering phase of two pions in infinite volume.  In the
derivation of this result, it is assumed that the physical box size is
large enough to avoid significantly altering the two pion interaction.
In this limit, 
the intrinsic finite size effects of each individual
pion are exponentially suppressed. The dominant contribution to the
finite size dependence of a two pion state is then simply given by the
interaction between the two pions due to the finite volume. Thus the
strength and nature, repulsive or attractive, of the interaction of
two pions will shift the energies of the otherwise free pions. This
shift then relates the energy eigenvalues to the scattering phase and
ultimately the scattering lengths of two pions.

One obstacle to the lattice determination of the pion-pion scattering
lengths is the presence of disconnected diagrams that render the
calculation of the $I=0$ channel computationally demanding.  On the
other hand, the simpler $I=2$ channel does not require such diagrams
and consequently many lattice groups have focused their efforts on
this case.  Furthermore, most calculations of the scattering lengths
to date have been carried out within the quenched
approximation~\cite{Ishizuka:2003nb,Aoki:2004wq,Aoki:2005uf,Sasaki:2007mt,Sasaki:2008sv,Sharpe:1992pp,Gupta:1993rn,Kuramashi:1993ka,Kuramashi:1993yu,Fukugita:1994na,Fukugita:1994ve,Aoki:1999pt,Fiebig:1999hs,Liu:2001zp,Aoki:2001hc,Liu:2001ss,Aoki:2002in,Aoki:2002sg,Aoki:2002ny,Juge:2003mr,Gattringer:2004wr,Li:2007ey}.
There have been only two previous calculations of $\atwo$ with
dynamical fermions.  The first such calculation was performed by
CP-PACS with $N_f=2$ tadpole-improved clover fermions at rather heavy
pion masses in the range $m_\pi=0.5~\gev$ to
$1.1~\gev$~\cite{Yamazaki:2004qb}.  However, it is doubtful that
\chipt{} at NLO, or any order, can be applied to such heavy pion
masses.  The other full QCD calculation was performed by NPLQCD with
domain-wall valence quarks on the $N_f=2+1$ asqtad-improved coarse
MILC ensembles with $m_\pi=290~\mev$ to $590~\mev$
\cite{Beane:2005rj,Beane:2007xs}. Mixed-action \chipt{} at NLO was
used to perform the chiral and continuum extrapolations. At the
physical pion mass, NPLQCD finds
\bd
m_\pi\atwo=-0.04330\,(42)\gap\textmd{and}\gap
\lpipi(\mu=\fphy)=6.2\,(1.2)\,, \ed
where $\lpipi(\mu)$ is a low energy constant (LEC) appearing
in the \chipt{} description of the quark mass dependence of the
scattering length.  As discussed later, $\lpipi(\mu)$ is
evaluated at $\mu=\fphy$, where $\fphy$ is the physical value of the
pion decay constant.

In this work we determine the s-wave $I=2$ pion-pion scattering length
and the corresponding $\lpipi$ by using the $N_f=2$
maximally twisted mass fermion ensembles from the European Twisted
Mass Collaboration (ETMC).  Our lightest pion mass is lighter than
those of the previous calculations and allows us to further probe the
chiral limit.  Due to the properties of twisted mass fermions at
maximal twist, our calculation is automatically accurate to $O(a^2)$
in the lattice spacing, $a$.  Additionally, we perform an explicit
check for large lattice artifacts with a single calculation at a finer
lattice spacing. As presented later, we find at the physical pion mass
\bd m_\pi \atwo=-0.04385\,(28)(38)
\gap\textmd{and}\gap
\lpipi(\mu=\fphy)=4.65\,(.85)(1.07)\,, \ed
which is in agreement with the above experimental measurement\new{s} and
phenomenological analysis as well as the previous lattice calculation.

\section{Method}

\subsection{L\"uscher's finite size method}

As mentioned in the introduction, L\"uscher's finite size method
relates the energy levels of two pion states in a finite volume to the
scattering phase in the infinite volume.  For the case of two pions
with zero total three-momentum, this method establishes a relationship
between the lowest energy eigenvalue $E_{\pi\pi}^I$ with a given
isospin $I$ in a finite box of size $L$ and the corresponding
scattering length $\aI$. For the $I=2$ channel, it is given in
\Ref{\cite{Luscher:1986pf}} as
\ba
\dE &=& \E - 2m_\pi \nonumber\\
&=&
- \frac{4\pi\atwo}{m_\pi L^3}
\left[ 1 + c_1\frac{\atwo}{L} + c_2\left(\frac{\atwo}{L}\right)^2 \right]
+ O(L^{-6})\,,
\label{eq:luscher}
\ea
where $c_1=-2.837297$ and $c_2=6.375183$ are numerical constants.
Thus the above result allows us to convert a lattice determination of
the energy shift, $\dE$, into a calculation of $\atwo$.

\subsection{Extraction of $\dE$}

To extract $\dE$, we construct the $\pi^+$ and $\pi^+\pi^+$
two-point correlation functions from the operators proposed in
\Ref{\cite{Fukugita:1994ve}},
\bd
C_{\pi}(t)= \langle(\pi^+)^\dagger(t+t_s)\pi^+(t_s)\rangle
\ed
and
\bd
C_{\pi\pi}(t) = \langle(\pi^+\pi^+)^\dagger(t+t_s)(\pi^+\pi^+)(t_s)\rangle\,.
\ed
Here $t_s$ is an arbitrary time slice, $\pi^+(t) = \sum_{\vec{x}}
(\bar{d}\gamma_5 u)(\vec{x},t)$ is an interpolating operator for the
$\pi^+$ meson with zero total three-momentum and $(\pi^+\pi^+)(t)$ is
an interpolating operator for the two pion state, again with zero total
three-momentum, given by
\bd
(\pi^+\pi^+)(t)=\pi^+(t+a)\pi^+(t)\,.
\ed
In order to avoid complications due to Fierz rearrangement of quark
lines as discussed in \Ref{\cite{Fukugita:1994ve}}, we use the
$\pi^+$ interpolating fields at time slices separated by one lattice
spacing.

From the large time behavior of $C_{\pi}(t)$ and $C_{\pi\pi}(t)$, it
is possible to extract the corresponding ground state energies as
follows,
\bd
C_{\pi}(t)\rightarrow A_{\pi}\exp(-m_{\pi}\,t)
\gap\textmd{and}\gap
C_{\pi\pi}(t)\rightarrow A_{\pi\pi}\exp(-E_{\pi\pi}^{I=2}\,t)\,,
\ed
where we assume that $t$ is large enough to neglect excited states but
still far enough from the boundaries to ignore boundary effects.
Furthermore, constructing the following ratio of correlation functions
we can determine $\dE$ directly as
\bd
\frac{C_{\pi\pi}(t)}{C_{\pi}^2(t)}\rightarrow \frac{A_{\pi\pi}}{A^2_{\pi}}\exp(-\dE\,t)
\ed
where $t$ satisfies the same requirements as before.  However, we use
anti-periodic boundary conditions for the quarks in the time direction
in order to match the sea quarks used in our calculation, and this leads
to a more complicated time dependence for $C_{\pi}$ and $C_{\pi\pi}$.

\subsection{Anti-periodic boundary conditions}

As mentioned above, in our calculation we employ anti-periodic
boundary conditions in the time direction for the fermions.  Using the
transfer matrix formalism, the time dependence of our correlation
functions is given by
\bd
\langle O^\dagger(t)O(0)\rangle=\textmd{Tr}\left(e^{-H(T-t)}O^\dagger(0)e^{-Ht}O(0)\right)/\,Z\,,
\ed
where the time-slice transfer matrix is $e^{-aH}$, the partition
function $Z$ is given by $Z=\textmd{Tr}(e^{-HT})$ where $T$ is the total
time extent of our lattice and $O(t)$
represents either $\pi^+(t)$ or $(\pi^+\pi^+)(t)$.  Inserting a
complete set of eigenstates of $H$ into the above equation yields
\begin{eqnarray*}
\langle
O^\dagger(t)O(0)\rangle&=&\sum_{m,n}|\langle
n|O|m\rangle|^2e^{-E_m(T-t)}e^{-E_nt}/\,Z\\
&=&\sum_{m,n}|\langle
n|O|m\rangle|^2e^{-(E_m+E_n)T/2}\cosh((E_m-E_n)(t-T/2))/\,Z\,.
\end{eqnarray*}
The terms in the above series are thermally suppressed by factors
of $e^{-E_m T}$ or $e^{-E_n T}$.  Only those terms with
$E_m=0$ or $E_n=0$ remain in the zero temperature,
$T\rightarrow\infty$, limit. However, the effects of the suppressed
contributions can still distort the behavior of correlation
functions for finite values of $T$, particularly in the large $t$
region.

This phenomenon does indeed occur in this work for the two pion
operator.  Intermediate states $\langle n|=\langle \pi^+|$ and
$\langle m|=\langle \pi^-|$ give a constant, in $t$, contribution
to $C_{\pi\pi}$,
\bd
|\langle \pi^+|\pi^+\pi^+|\pi^-\rangle|^2e^{-m_\pi T}/Z\,.
\ed
This is comparable to the standard contribution,
\bd
|\langle \pi^+\pi^+|\pi^+\pi^+|\Omega\rangle|^2e^{-E_{\pi\pi}^{I=2}T/2}\cosh(E_{\pi\pi}^{I=2}(t-T/2))/Z\,,
\ed
when $t$ approaches $T/2$.  To be precise, for large enough volumes
$E^{I=2}_{\pi\pi} = 2 m_\pi + \dE \approx 2 m_\pi$, and hence
these two contributions to $C_{\pi\pi}$, $e^{-m_\pi T}$ and
$e^{-E_{\pi\pi}^{I=2}T/2}\cosh(E_{\pi\pi}(t-T/2))$ are in fact nearly equal for $t=T/2$.
Additionally, the factor $C_\pi(t)^2$ has similar problems.  The correlator
$C_\pi(t)$ itself has a simple spectral representation.  However,
the square is more complicated and also contains a constant, in $t$, contribution
as well.

To eliminate these contaminations, we use the derivative
method~\cite{Umeda:2007hy} and define a modified ratio, $R(t)$, in the
following way
\be
\label{eq:lattice_ratio}
R(t+a/2)=\frac{C_{\pi\pi}(t)-C_{\pi\pi}(t+a)}{C_{\pi}^2(t)-C_{\pi}^2(t+a)}\,.
\ee
The asymptotic form for $R(t)$, ignoring terms suppressed relative to
the leading contribution, is
\be
\label{eq:asymptotic_ratio}
R(t+a/2)=A_R\left(\cosh(\dE\,t^\prime)+\sinh(\dE\,t^\prime)\,\coth(2m_\pi t^\prime)\right)
\ee
where $A_R$ is a combination of amplitudes in $C_\pi$ and $C_{\pi\pi}$
and $t^\prime = t+a/2-T/2$.  Since $m_\pi$ is the most accurately
calculated component of our calculation, $R(t)$ provides a nearly
direct determination of $\dE$ and cleanly eliminates the unwanted
thermal contributions that spoil the simple ratio given earlier.

\section{Lattice Calculation}

\subsection{Twisted mass fermions}

In this work we use the two flavor maximally twisted mass fermion
configurations from ETMC.  Using twisted mass fermions at maximal
twist ensures that physical observables are automatically accurate
to $O(a^2)$ in the lattice spacing~\cite{Frezzotti:2003ni}.  Most of
the results presented here are from a sequence of ensembles with a
lattice spacing of $a=0.086~\fm$ and a box size of $L=2.1~\fm$.  The
pion masses range from $m_\pi=270~\mev$ to $485~\mev$.  For the
lower pion masses the volume is increased to $L=2.7~\fm$, and there
is one calculation using a finer lattice spacing of $a=0.067~\fm$.
The parameters relevant to this calculation are given in
\Tab{\ref{tab:ensemble}}, and further details can be found in
\Refs{\cite{Boucaud:2007uk,Dimopoulos:2008sy,Urbach:2007rt}}.

\begin{table}[tb]
\begin{center}
\begin{tabular}{|c|c|c|c|c|c|c|c|}
\hline
$\beta$ & $a\mu$ & $L/a$
& $m_\pi$ &$m_{\pi}/f_{\pi}$ & $N$ & $a\dE\cdot 10^{3}$ & $m_\pi \atwo$ \\\hline
3.90  & 0.0100 & $24$ & 485 & 2.77(2) & 479 & 7.23(59)(41) & -0.297(20)(16) \\\hline
3.90  & 0.0085 & $24$ & 448 & 2.61(1) & 487 & 7.66(65)(33) & -0.269(17)(10) \\\hline
3.90  & 0.0064 & $24$ & 391 & 2.40(1) & 553 & 9.6(1.3)(.6) & -0.252(22)(13) \\\hline
3.90  & 0.0040 & $32$ & 309 & 2.02(1) & 490 & 3.96(36)(22) & -0.165(14)(08) \\\hline
3.90  & 0.0030 & $32$ & 270 & 1.85(1) & 562 & 4.05(42)(21) & -0.130(12)(06) \\\hline
4.05 & 0.0030 & $32$ & 307 & 2.08(2) & 375 & 7.1(1.2)(.9) & -0.171(18)(22) \\\hline
\end{tabular}
\end{center}
\vspace{10pt}
\caption{Ensembles used in this work.  Only dimensionless quantities are
needed in this calculation, but for guidance we give the value of
$m_\pi$ rounded to the nearest $\mev$ for each ensemble indicated
by $\beta$, $a\mu$ and $L/a$. We also list the ratio $m_\pi/f_\pi$,
the number, $N$, of configurations used, the energy shift $a\dE$ and the
scattering length $m_\pi \atwo$.  The first uncertainty is statistical and, when present,
the second one is systematic.}
\label{tab:ensemble}
\end{table}

\subsection{Stochastic sources}

For the calculation of pion correlation functions, it is known that
the stochastic source method is more efficient than the point source
method.  Therefore, in the present work, we employ $Z_4$ stochastic
noise with two noise sources generated on each source time slice.
Since we place the source on two time slices for the $\pi^+\pi^+$
correlation function, $t_s$ and $t_s+a$, we therefore perform four
inversions for each configuration. We remark that we also use the
one-end trick in this work for the evaluation of correlation
functions~\cite{Foster:1998vw,McNeile:2006bz,Boucaud:2008xu} leading
to a further improvement in the signal-to-noise ratio. Additionally,
the source time slices, $t_s$, are chosen randomly to reduce the
autocorrelation between consecutive trajectories.

\section{Results}

\subsection{Calculation of $m_\pi \atwo$}

\begin{figure}[tb]
\begin{center}
\hspace{-15pt}\includegraphics[width=280pt,angle=\plotangle]{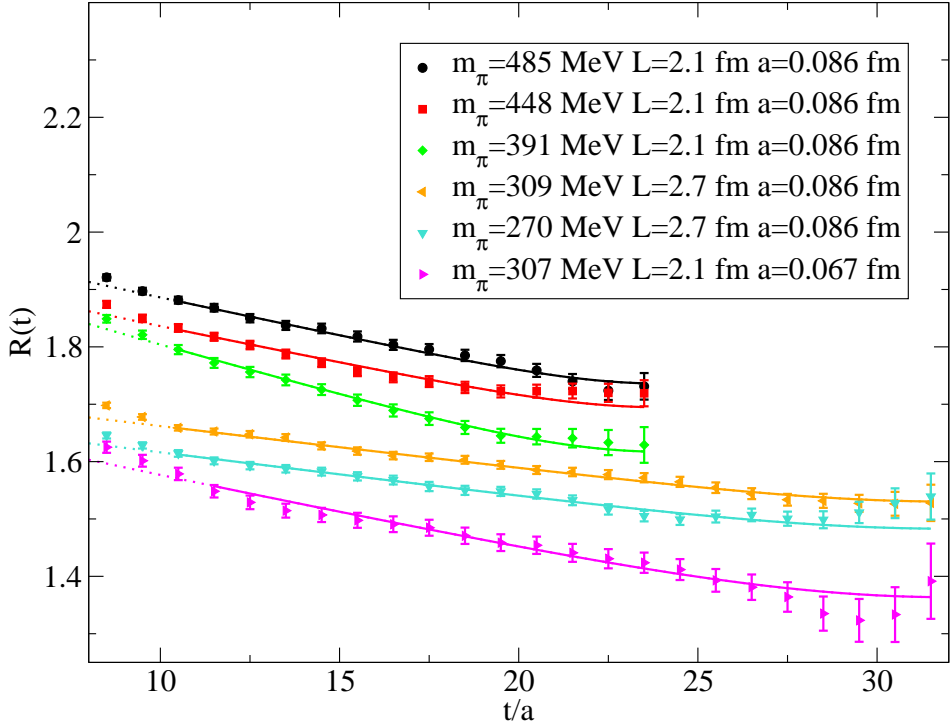}
\end{center}
\caption{The ratio $R(t)$ as a function of $t$.  The solid lines are
correlated fits to \Eq{\ref{eq:asymptotic_ratio}}, from which the energy
shifts $a\dE$ are extracted. The ensembles have been shifted vertically to facilitate
easier comparison.}
\label{fig:ratio}
\end{figure}
In \Fig{\ref{fig:ratio}} we show our lattice results for $R(t)$,
defined in \Eq{\ref{eq:lattice_ratio}}, as a function of the time
$t$ together with a correlated fit to the asymptotic form given in
\Eq{\ref{eq:asymptotic_ratio}}.  All the ensembles shown in
\Fig{\ref{fig:ratio}} visibly agree with the corresponding fit and
lead to reasonable values of $\chi^2$ per degree of freedom
($\mathrm{dof}$), where $\chi^2$ is the correlated figure-of-merit
function.  To further verify these fits, we examined several
possible sources of systematic error.  First, the ratio could suffer
from bias at large $t$, so we examined the jackknife estimate of
bias but found it to be significantly smaller than the errors for
all the ensembles.  Second, we considered the possibility that we
underestimated the errors due to autocorrelations.  However, both
the gamma method~\cite{Wolff:2003sm} and standard binning showed no
significant signs of autocorrelation for $R(t)$ in any of the
ensembles.  The possibility of $\pi^0$ mixing, due to the breaking
of parity at non-zero lattice spacing for twisted mass fermions, is
considered in \Sect{\ref{sect:pio}}.  But as described there in more
detail, we find no statistically significant indications of the
$\pi^0$ contributions.

There is one further possible systematic error due to the
contributions from excited states in the small $t$ region or from
unphysical $\pi^0$ states in the large $t$ region.  To ensure that the
fits for these ensembles are safe from such effects, we study the
systematic errors caused by choosing a fitting window in which to
match to the asymptotic form for $R(t)$.  First we ensure that the
results exhibit clear plateaus when we increase the minimum $t$ or
decrease the maximum $t$ used in the fits.  However, to provide a
quantitative estimate of the systematic error, we perform the
following distribution method.  We collect the results for $a\dE$ from
all fitting intervals with $\chi^2/\mathrm{dof}<2$.  This includes
varying both the minimum and maximum time extent for the fitting range
and results in $30$ to $60$ values of $a\dE$ for each ensemble.  We
then make the distribution of these selected results and choose the
median of this distribution for the central value.  Then we take the
central, and symmetric about the median, $68\%$ region of the
distribution to define the systematic error.  Finally, we use the
jackknife method to determine the statistical error on the central
values.  This method is also applied to $m_\pi \atwo$, and the results
for $a\dE$ and $m_\pi \atwo$ are given in \Tab{\ref{tab:ensemble}}.
As shown in this table, the resulting estimates of the systematic
errors are typically smaller than the corresponding statistical
errors, and are at worst of the same order as the statistical errors.
Since the distribution method used to estimate the systematic errors
is itself subject to statistical errors, this is precisely what is
expected if there are no substantial systematic effects.  However,
since the final statistical precision for the value of $m_\pi \atwo$
at the physical limit turns out to be quite small, we decided, in
order to avoid underestimating our final error, to carefully propagate
these systematic errors through to the final result as described later
in \Sect{\ref{sect:chiral_extrapolation}}.

\subsection{$\pi^0$ contamination}
\label{sect:pio}

Twisted mass fermions violate parity and isospin at non-zero values of
the lattice spacing.  Therefore the spectral representation of the
$\pi^+$ and $\pi^+\pi^+$ correlators can admit states that would not
be present in the continuum limit.  In particular, unphysical
contributions from the $\pi^0$, which has a mass $m_{\pi^0}$
different, and smaller, than the mass $m_\pi$ of the $\pi^\pm$, may
enter the $C_\pi$ and $C_{\pi\pi}$ correlators in several
ways~\cite{Boucaud:2008xu}.  Furthermore, these effects are believed
to be more noticable in pion-pion scattering, so the successful
calculation of all three isospin channels, $I=0$, $1$ and $2$, would
test the twisted mass formulation of lattice QCD.

The $\pi^0$ can enter the $C_\pi$ correlator through intermediate
states of the form $\langle\pi^+|\pi^+|\pi^0\rangle$ and
$\langle\pi^+\pi^0|\pi^+|\Omega\rangle$.  The former contribution is
thermally suppressed by a factor of $e^{-m_{\pi^0} T}$, however it
leads to a time dependence with an energy of $m_\pi-m_{\pi^0}$ that is
lighter than the usually expected $m_\pi$ ground state.  The second
contribution is not thermally suppressed but corresponds to the first
excited state with energy $E_{\pi^+\pi^0}\approx m_\pi + m_{\pi^0}$.
This is lighter than the first physical excited state with energy near
$3 m_\pi$.  Similarly, $C_{\pi\pi}$ contains unphysical contributions
from $\langle\pi^+\pi^+|\pi^+\pi^+|\pi^0\rangle$ and
$\langle\pi^+\pi^+\pi^0|\pi^+\pi^+|\Omega\rangle$.  Again there is an
additional light state that is thermally suppressed by
$e^{-m_{\pi^0}T}$ but has an energy of
$E_{\pi^+\pi^+}-m_{\pi^0}\approx 2 m_\pi - m_{\pi^0}$ that is lower
than the physical ground state near $2m_\pi$, and the first excited
state is lowered to $E_{\pi^+\pi^+\pi^0}\approx 2m_\pi + m_{\pi^0}$
rather than the expected energy of approximately
$2\sqrt{m_\pi^2+(2\pi/L)^2}$.

The parity violating matrix elements responsible for these effects are
$O(a)$ in the lattice spacing, even at maximal twist, however the
matrix elements appear squared in the correlators.  Therefore these
unphysical states make an $O(a^2)$ contribution.  The question,
however, is not about the scaling in the lattice spacing, but about
the size of this contribution at the lattice spacings used in this
work.  A detailed discussion of this issue can be found in
\Ref{\cite{Dimopoulos:2009qv}}.  Here, our focus is more practical.
We want to ensure that the scattering lengths calculated in this work
are not significantly distorted due to these effects.

First, the naive estimate for the suppression factor for the
additional light contributions, $m_\pi-m_{\pi^0}$ in $C_{\pi}$ and
$E_{\pi^+\pi^+}-m_{\pi^0}$ in $C_{\pi\pi}$, is
$(a\Lambda_\mathrm{QCD})^2 e^{-m_{\pi^0} T}$.  The value of
$m_{\pi^0}$ is difficult to calculate precisely, but it is clear from
\Ref{\cite{Dimopoulos:2009qv}} that $m_{\pi^0}$ is never more than
$20\%$ lighter than $m_\pi$ for the ensembles in this work.  Therefore
we will simply use $m_\pi$ and a value of
$\Lambda_\mathrm{QCD}=250~\mev$ to set the order of magnitude for
these suppression factors.  We find that for the ensembles used here,
the largest value of $(a\Lambda_\mathrm{QCD})^2 e^{-m_{\pi^0} T}$ is
$9\cdot 10^{-6}$ for the $\beta=4.05$, $a\mu=0.0030$ ensemble in
\Tab{\ref{tab:ensemble}}.  Using the actual value of $m_{\pi^0}$ from
\cite{Dimopoulos:2009qv} raises this to $2\cdot 10^{-5}$.  This value
is small, but it is not too far beyond the statistical precision of
the correlators used to calculate $a\dE$, hence we must carefully
check for these contributions.

Second, there are the additional states that are only suppressed by
$(a\Lambda_\mathrm{QCD})^2$.  However, these states are heavier than
the physical state and hence would occur in the correlators as excited
states.  The naive suppression factors are $1\cdot 10^{-2}$ and
$7\cdot 10^{-3}$ for $a=0.086~\fm$ and $0.067~\fm$ respectively.
These simple estimates are larger than for the other states, however
these contributions are also more strongly suppressed by their own
energies.

In the light of these arguments, we made a significant effort to
attempt to find such effects anyway.  We tried fitting the
individual $C_\pi(t)$ and $C_{\pi\pi}(t)$ correlators as well as the
ratio $R(t)$ to various functional forms including the physical
state and both the additional heavier and lighter states, just the
lighter state or just the heavier state.  We fit the most general
forms, keeping all energies as free parameters, and
additionally constrained forms, in which we constrained $m_{\pi^0}$
based on known values. And we also explored several minimization
methods. The net result was that one could indeed lower the $\chi^2$
value for each fit, but the $\chi^2$ per
degree of freedom still increased, indicating no statistically
significant contribution from the unwanted $\pi^0$ states.

However, we must offer a few words of caution.  While we could not
find any compelling evidence for these contributions, we of course can
not rule out their presence at a level beneath our statistical
resolution.  We should further note that there are visible excited
states in the correlators.  However, the accuracy of the correlators
for the ensembles studied here does not allow us to distinguish the
physical excited states, near $3m_\pi$ for $C_\pi$ and
$2\sqrt{m_\pi^2+(2\pi/L)^2}$ for $C_{\pi\pi}$, from the unphysical
excited states, near $m_\pi+m_{\pi^0}\approx 2 m_\pi$ for $C_\pi$ and
$2m_\pi+m_{\pi^0}\approx 3 m_\pi$ for $C_{\pi\pi}$.  The extensive
study of systematic errors due to the fitting range discussed in the
previous section was partially motivated by these issues.  It provides
the quantitative statement that these effects do not rise to the level
of our statistical precision and gives an estimate of the systematic
error.

Additionally, there are two reasons that these contributions may be
smaller than anticipated.  First, the unphysical contributions
correspond to scattering states that may be suppressed by a power of
the volume.  Second, the construction of $R(t)$ in
\Eq{\ref{eq:lattice_ratio}} forms a discrete approximation to the
ratio of derivatives of $C_{\pi\pi}$ and $C_\pi^2$ and may further
suppress the nearly constant light state contributions. Finally, these
effects are anticipated to be more substantial for the other isospin
channels, hence a detailed understanding of the $\pi^0$ contributions
to pion-pion scattering will have to await our ongoing calculations in
the $I=1$~\cite{Feng} channel and our planned work for $I=0$.

\subsection{Finite volume effects}

The dominant finite size effect in this calculation is, of course,
the shift in $\dE$ due to the interactions of two pions in a finite
volume.  Additionally, there are the exponentially small, as opposed
to the merely power suppressed, finite volume corrections to $I=2$
pion-pion scattering that have been determined for scattering near
threshold in \Ref{\cite{Bedaque:2006yi}}.  The resulting finite size
corrections for the scattering length are given there as,
\bd
(m_\pi \atwo)_{L}=(m_\pi \atwo)_{\infty}+\Delta_{FV}
\ed
where
\begin{eqnarray*}
\label{eq:finite_volume} \Delta_{FV}&=&-\frac{m_\pi^2}{8\pi
f_\pi^2}\left\{\frac{m_\pi^2}{f_\pi^2}\frac{\partial}{\partial
m_\pi^2} i\Delta\mathcal{I}(m_\pi)+\frac{2m_\pi^2}{f_\pi^2}i\Delta
\mathcal{J}_{exp}(4m_\pi^2)\right\}  \nonumber\\
&=&\frac{1}{2^{13/2}\pi^{5/2}}\left(\frac{m_\pi}{f_\pi}\right)^4\sum_{|{\bf{n}}|\neq0}
\frac{e^{-|{\bf{n}}|m_\pi L}}{\sqrt{|{\bf{n}}|m_\pi L}}
\left\{1-\frac{17}{8}\frac{1}{|{\bf{n}}|m_\pi L}+O\left(L^{-2}\right)\right\}\,.
\end{eqnarray*}
Using the above result, we calculate the corrections to $m_\pi\atwo$.
Compared to the statistical errors, the
finite volume corrections are negligible.  To be precise, they are
never more than 6\% of the corresponding statistical error and are
hence ignored in the following analysis.

\old{There is a second finite size effect of a technical nature.  The
effective range approximation is used in L\"uscher's method to relate
the scattering phase to the scattering length.}
\new{
There is a second finite size effect originating from the
effective range approximantion, $p\tan^{-1}\delta(p)=1/\atwo+\frac{1}{2}r_\mathrm{eff}p^2$,
which is used
to relate the scattering phase $\delta(p)$ at vanishingly small momentum $p$ to the scattering length.
The dependence on the effective range $r_\mathrm{eff}$ is very small and gives rise to the corrections
at $O(L^{-6})$ in \Eq{\ref{eq:luscher}}.
}
As argued in \Ref{\cite{Beane:2007xs}},
assuming that the effective range is \old{of the order of}\new{at most twice} the scattering length,
this correction can be estimated using the measured values of $m_\pi$ and
$\dE$.  Using the result given in \Ref{\cite{Beane:2007xs}}, we calculate
this correction and find that it is never more than 9\% of the corresponding
statistical error of $m_\pi\atwo$.  Hence, this finite size effect is also
sufficiently small to be ignored as well.

\subsection{Lattice artifacts}

Most of the calculations presented here use a single lattice spacing
of $0.086~\fm$, but we have also performed an additional calculation
of $\dE$ and $m_\pi\atwo$ at a second lattice spacing of $0.067~\fm$
and at a pion mass of $307~\mev$.  This pion mass lies very close to
that of the $a=0.086~\fm$, $m_\pi=309~\mev$ point.  The physical
volumes of these two ensembles differ, so the values of $\dE$ can
not be directly compared.  However, assuming that L\"uscher's method
correctly accounts for the finite volume dependence of $\dE$ for
these two ensembles, we can compare $m_\pi \atwo$ for the two
lattice spacings, and indeed we do find statistical agreement
between the two ensembles as indicated in \Tab{\ref{tab:ensemble}}.
Furthermore, as described in the next section, we note that the
expected $O(a^2)$ corrections from maximally twisted mass lattice
QCD are actually weakened to $O(m_\pi^2 a^2)$ for the $I=2$,
$I_3=\pm2$ channel as shown using twisted mass
{\chipt}~\cite{Buchoff:2008hh}, thus suggesting further that the
lattice spacing dependence of $m_\pi\atwo$ is mild for the
calculations in this work.

\subsection{Chiral extrapolation}
\label{sect:chiral_extrapolation}

The pion-pion scattering lengths have recently been calculated in
twisted mass {\chipt}~\cite{Buchoff:2008hh}.  This is an expansion
of twisted mass lattice QCD in both the quark masses and the lattice
spacing.  There it is shown that at NLO the lattice spacing
corrections to the $I=2$, $I_3=\pm2$ scattering lengths are
proportional to $\cos(\omega)$, where $\omega$ is the twist angle.
Thus at maximal twist, $\omega = \pi/2$, the explicit discretization
errors vanish exactly, and the scattering length can be simply
represented by the continuum NLO {\chipt}
formula~\cite{Gasser:1983yg,Gasser:1983kx}.

As suggested in \Refs{\cite{Beane:2005rj,Beane:2007xs}}, we perform
the chiral extrapolation of $m_\pi \atwo$ in terms of $m_\pi/f_\pi$
instead of $m_\pi$.  Additionally, the {\chipt} renormalization
scale is fixed as $\mu=\fphy$.  The resulting NLO expression is then
\be
\label{eq:fit}
m_\pi \atwo
 = -\frac{m_\pi^2}{8\pi f_\pi^2}\left\{1+\frac{m_\pi^2}{16\pi^2 f^2_\pi}
\left[3\ln \frac{m_\pi^2}{f_\pi^2}-1-\lpipi(\mu=\fphy)\right]\right\}\,,
\ee
where $\lpipi(\mu)$ is related to the Gasser-Leutwyler
coefficients $\bar{l}_i$ as~\cite{Bijnens:1997vq}
\bd
\lpipi(\mu)=\frac{8}{3}\bar{l}_1+\frac{16}{3}\bar{l}_2-\bar{l}_3-4\bar{l}_4+
3\ln\frac{m_{\pi,\mathrm{phy}}^2}{{\mu^2}}\,.
\ed
It is important to note that extrapolating in $m_\pi/f_\pi$ instead of
simply $m_\pi$ does indeed change the expression for $m_\pi \atwo$ but
only at the next-to-next-to-leading order (NNLO).  The advantage of
this form is that $m_\pi/f_\pi$ is calculated directly on the lattice
with small errors and the chiral extrapolation does not require fixing
a physical value for the lattice spacing.

\begin{figure}[htb]
\begin{center}
\hspace{-15pt}\includegraphics[width=280pt,angle=\plotangle]{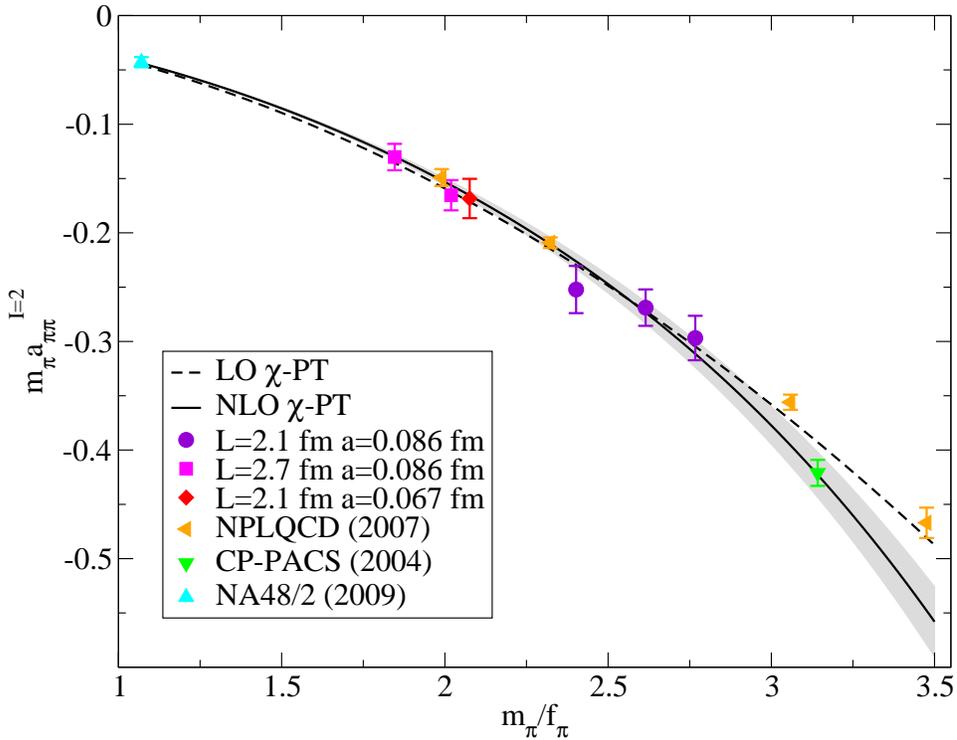}
\end{center}
\caption{Chiral extrapolation for the I=2 pion-pion scattering
length. The results in this work are shown together with the lattice
calculations of NPLQCD~\cite{Beane:2005rj,Beane:2007xs} and
CP-PACS~\cite{Yamazaki:2004qb} and the \old{experimental data from E865
at BNL~\cite{Pislak:2003sv}}\new{direct measurement from NA48/2 at CERN~\cite{NA48}}.} \label{fig:fit}
\end{figure}

We now fit our lattice results for $m_\pi \atwo$ from
\Tab{\ref{tab:ensemble}} to the functional form in \Eq{\ref{eq:fit}}
in order to extrapolate $m_\pi \atwo$ to the physical point and also
extract the low energy constant $\lpipi(\mu=\fphy)$. The
calculated values for the scattering length and the resulting {\chipt}
fit curve are shown in \Fig{\ref{fig:fit}}.
In the same figure, we also provide a comparison to the lattice
results of NPLQCD~\cite{Beane:2005rj,Beane:2007xs} and
CP-PACS~\cite{Yamazaki:2004qb} and the \old{experimental data from E865
at BNL~\cite{Pislak:2003sv}}\new{direct measurement from NA48/2 at
CERN~\cite{NA48}}.  We find general agreement between our
calculation and the results of NPLQCD at similar pion masses.
\old{Additionally, we find agreement with the experimental determination
of $m_\pi\atwo$.}\new{
In particular, the agreement between our results 
and NPLQCD suggests that the effect of the missing strange quark
in our current calculation is small.  Additionally, 
the ongoing effort of
ETMC to include the dynamical effects of both the strange and charm
quark~\cite{Chiarappa:2006ae,Baron:2008xa} will allow us to directly address this issue.
}

\begin{figure}[htb]
\begin{center}
\hspace{-15pt}\includegraphics[width=280pt,angle=\plotangle]{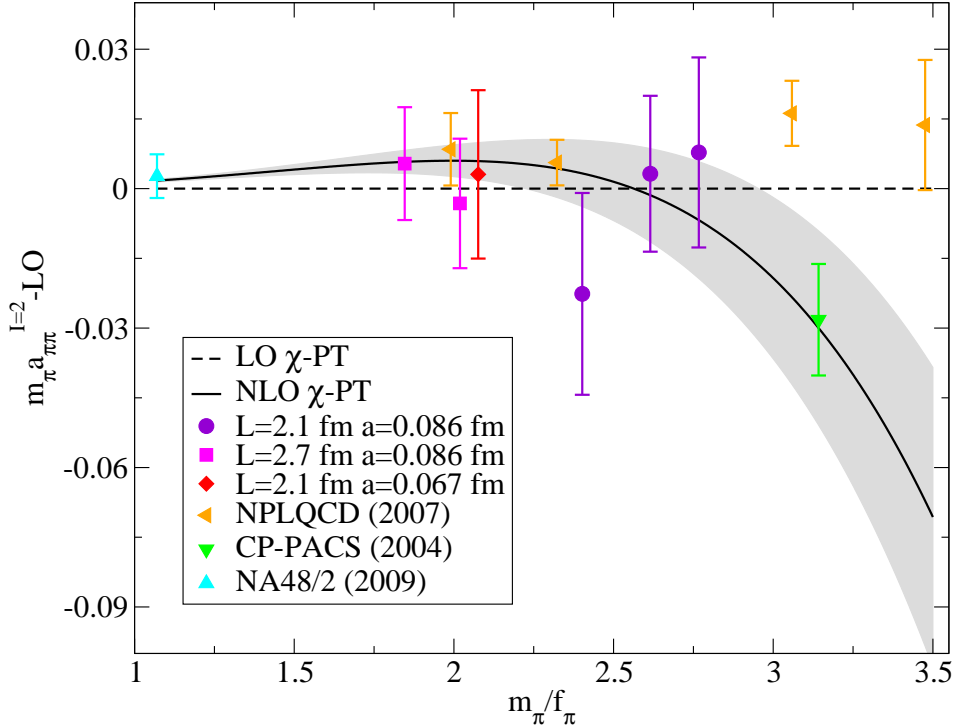}
\end{center}
\caption{Difference between the lattice calculation of the
scattering lengths and the LO {\chipt} prediction. The scattering
lengths agree statistically with the LO {\chipt} prediction for
$m_\pi =270~\mev$ to $485~\mev$.} \label{fig:fit1}
\end{figure}

To highlight the impact of the NLO terms in the {\chipt} description
of the pion mass dependence of $m_\pi\atwo$ and to understand the role
of yet higher order terms,
we show the difference between the lattice calculations of the scattering
length and the LO {\chipt} prediction in \Fig{\ref{fig:fit1}}.
We find that the scattering lengths statistically agree with the LO
{\chipt} result for all lattice calculations with $m_\pi <
500~\mev$.  Accordingly, the NLO {\chipt} functional form provides a
reasonable description of the lattice results in the same region of
$m_\pi$.  As a further check, we fit our calculations to the NNLO
form for $m_\pi\atwo$~\cite{Colangelo:2001df,Bijnens:1997vq} and
found $m_\pi\atwo=-0.041\,(12)$ at the physical point.  The
statistical error is large, as one would expect given that our
results already agree statistically with the LO {\chipt} form, but
the resulting NNLO extrapolation of $m_\pi\atwo$ does agree with the
NLO fit. Given the size of the statistical errors, we are unable to
make any meaningful estimate of the NNLO LECs, however, the
effects from truncating the {\chipt} series to NLO is included in our
estimate of systematic errors.

The systematic error on the extrapolated value of $m_\pi \atwo$ and
$\lpipi$ has several components.  First, the systematic errors
of the $m_\pi \atwo$ that we obtain for each ensemble are propagated
through the chiral extrapolation. This is accomplished by again
collecting all fit ranges for each ensemble with
$\chi^2/\mathrm{dof}<2$ as earlier.  This gives approximately
$10^{10}$ {\chipt} fits from which we randomly choose $2000$ to
sample the distribution of the extrapolated values of $m_\pi \atwo$.
As for the individual $m_\pi \atwo$, we use the distribution method
to determine an estimate of the systematic error due to the fit
ranges from each ensemble. The second systematic uncertainty arises
from the chiral fit itself. This is estimated by taking the
difference in the extrapolated values from the NLO {\chipt} fit to
all six and just the lightest five ensembles.  Finally, the extrapolation
to the physical point requires the experimental value for $m_\pi/f_\pi$.
The experimental error on this quantity introduces an error that is
nearly 50\% of the corresponding statistical error and hence is
also included.
All three effects are
added in quadrature to form the total estimated systematic error.
Using the latest PDG~\cite{Amsler:2008zzb} values of
$m_{\pi^+}=139.5702(4)~\mev$ and $f_{\pi^+}=130.4(2)~\mev$ to
determine the physical limit, we obtain the final result
\bd
m_\pi \atwo=-0.04385\,(28)(38) \gap\textmd{and}\gap
\lpipi(\mu=\fphy)=4.65\,(.85)(1.07)\,.
\ed
This agrees with the previously mentioned results:\ the lattice
calculation from NPLQCD~\cite{Beane:2005rj,Beane:2007xs}, the
so-called CGL analysis~\cite{Colangelo:2000jc,Colangelo:2001df} and
the \old{E865 measurement~\cite{Pislak:2003sv}}\new{E865~\cite{Pislak:2003sv}
and NA48/2~\cite{NA48} measurements} and represents agreement
among the experimental and theoretical determinations of $m_\pi\atwo$
at the 1\% level.
\old{In particular, the agreement between our results 
and NPLQCD suggests that the effect of the missing strange quark
in our current calculation is small.  Additionally, 
the ongoing effort of
ETMC to include the dynamical effects of both the strange and charm
quark~\cite{Chiarappa:2006ae,Baron:2008xa} will allow us to directly address this issue.}


\new{
This accuracy of 1\% must be understood as a
  combined theoretical effort from lattice QCD and chiral perturbation
  theory.
The quantity $m_\pi\atwo$, as well as $\atwo$
  itself, vanishes in the chiral limit.  This significantly constrains
  the chiral extrapolation of $m_\pi\atwo$.  In particular,
  $m_\pi\atwo$ is uniquely predicted in terms of $m_\pi/f_\pi$ at LO
  and depends only on one unknown constant, $\lpipi$, at NLO.  This
  makes the chiral extrapolation of lattice results particularly
  accurate.  Thus the 6\% to 11\% accurate results for the lattice calculation
  extrapolate to a 1\% accurate determination of $m_\pi\atwo$.  The final
result differs from the LO {\chipt} prediction by only a few percent, but
to illustrate the power of combining lattice QCD and {\chipt}, we note
that the difference between $m_\pi\atwo$ and the LO result is $0.00173\,(47)$,
which represents a $3.7\sigma$ shift that is due to the inclusion of
the NLO effects as determined by matching directly to lattice QCD.
}

\new{Chiral perturbation theory plays a strong role in obtaining
$m_\pi\atwo$ accurately, but it alone can not determine $\lpipi$.
Only in combination with the lattice results can we calculate
 $\lpipi =4.65\,(1.37)$.  This calculation is just under 30\%
accurate, however, as we explain shortly, this easily exceeds the experimental determinations
of this quantity.
The experimental and
  phenomenological results for $m_\pi\atwo$ at the physical point can
  be converted into a result for $\lpipi$ at NLO.  A simple analysis
  gives the following for $\lpipi$:\ $3.0\pm 3.1$ (CGL), $0.0\pm 10.3$ (E865 with {\chipt}) and
  $3.0\pm 2.8$ (NA48/2 with {\chipt}).
This comparison demonstrates the particular
  advantage of lattice calculations arising from the ability to vary
  the underlying quark masses of QCD.}

\section{Conclusion}

We have calculated the s-wave pion-pion scattering length in the
isospin $I=2$ channel using the two-flavor maximally twisted mass
lattice QCD configurations from ETMC. The pion masses ranged from
$270~\mev$ to $485~\mev$ and the lattice spacing was $a=0.086~\fm$.
\old{A
second lattice spacing of $a=0.067~\fm$ was used to demonstrate the
absence of large lattice artifacts. Additionally, the calculation is
accurate to $O(a^2)$ due to the properties of maximally twisted mass
fermions. Furthermore, discretization errors vanish from the $I=2$,
$I_3=\pm 2$ channel at NLO, as shown by twisted mass {\chipt}, hence we
extrapolated our results for the scattering length to the physical
limit using continuum \chipt{} at NLO.}\new{A
second lattice spacing of $a=0.067~\fm$ was used to demonstrate the
absence of large lattice artifacts.   This is only a single check,
 but when combined with the fact that the calculation is 
accurate to $O(a^2)$ due to the properties of maximally twisted mass
fermions, it suggests that the lattice spacing dependence is mild.
 Furthermore, discretization errors vanish from the $I=2$,
$I_3=\pm 2$ channel at NLO, as shown by twisted mass {\chipt}, hence we
extrapolated our results for the scattering length to the physical
limit using continuum \chipt{} at NLO.}
We investigated various
systematic effects, and we found for the scattering length at the
physical point $m_\pi \atwo=-0.04385\,(28)(38)$ and for the low energy
constant $\lpipi(\mu=\fphy)=4.65\,(.85)(1.07)$.  These
results are in good agreement with the previous lattice calculation
from NPLQCD, the experimental determination\new{s} from E865 at BNL
\new{and from NA48/2 at CERN}
 and the
CGL analysis using various theoretical and experimental inputs.

\section{Acknowledgment}

This work was supported by the DFG project Mu 757/13 and the DFG
Sonderforschungsbereich / Transregio SFB/TR9-03. We thank \new{B.~Bloch-Devaux, }G.~Herdoiza,
A.~Shindler, C.~Urbach and M.~Wagner for valuable suggestions and
assistance. X.~Feng would like to thank A.~Walker-Loud for helpful
correspondence regarding \chipt. The computer time for this project
was made available to us by the John von Neumann Institute for
Computing on the JUMP and JUGENE systems in J\"ulich. We also thank
the staff of the computer center in Zeuthen for their technical
support.

\bibliography{pion_scattering}
\bibliographystyle{unsrt_nt}

\end{document}